\documentclass[conference,compsoc]{IEEEtran}

\ifCLASSINFOpdf
\else
\fi

\usepackage{amsmath,amssymb}
\usepackage{booktabs}
\usepackage{cite}
\usepackage{graphicx}
\usepackage{multirow}
\usepackage{array}
\usepackage{xcolor}
\usepackage{url}
\usepackage{balance}

\title{Low-Cost Device-Free Material and Obstruction Identification in IoT Systems \\Using Commodity Bluetooth Low Energy RSSI }

\author{
\IEEEauthorblockN{Aditi Tripathi}
\IEEEauthorblockA{
Dept. of ECE\\
Banasthali Vidyapith, Rajasthan, India\\
Email: itsaditi2005@gmail.com
}
\and
\IEEEauthorblockN{Janhavi Tiwari}
\IEEEauthorblockA{
Dept. of ECE\\
MMMUT, Gorakhpur, India\\
Email: janhavitiwari009@gmail.com
}
\and
\IEEEauthorblockN{Sachin Kadam}
\IEEEauthorblockA{
Dept. of ECE\\
MNNIT Allahabad, Prayagraj, India\\
Email: sachink@mnnit.ac.in
}
}

\begin{document}
\maketitle

\begin{abstract}
Internet of Things (IoT) devices equipped with Bluetooth Low Energy (BLE) radios provide received signal strength indicator (RSSI) measurements for link management; however, these readily available measurements can also capture signal variations caused by materials obstructing the propagation path. This paper investigates material identification using two commodity Nordic nRF5340 development kits without an additional RF sensing front end. First, the distance-dependent RSSI baseline is characterized for the transmitter-receiver through experiments and demonstrates the need for distance-aware calibration. Next, RSSI traces are collected for five well-known materials as obstructions: namely, wood, ceramic, an empty plastic bottle, glass, and a human body. The traces are characterized using their mean, standard deviation, and transient deviation from a local baseline, and then the path-loss exponent is estimated. A lightweight threshold-based classifier is designed using these estimates, and thus the designed classifier is shown to correctly identify the human body, glass, and wood in the reported unknown trials, while materials like paper and brass are mapped to the most similar trained classes, namely a plastic bottle and ceramic. The results demonstrate the feasibility of coarse material identification using commodity BLE RSSI. 
\end{abstract}


\section{Introduction}
Wireless transceivers are increasingly used not only for communication but also as sensors of their surrounding environment. Propagating radio waves interact with objects through absorption, reflection, diffraction, and scattering. Consequently, the received waveform contains information about both the communication channel and the objects intersecting it. Recent wireless material identification systems have exploited these effects using technologies such as millimeter-wave radar, ultra-wideband (UWB), Wi-Fi channel state information (CSI), and radio-frequency identification (RFID) \cite{chen2025survey}. Although these systems can achieve rich sensing capabilities, they often require specialized transceivers, wide bandwidths, antenna arrays, or access to physical-layer measurements that are not available on low-cost embedded devices.

BLE offers a different design point. It is widely integrated into resource-constrained Internet-of-Things (IoT) platforms, and its radio stack already reports the received signal strength indicator (RSSI) for received packets. RSSI is a coarse scalar measurement that is sensitive to distance, antenna orientation, multipath propagation, radio channel, and temporal interference \cite{ramirez2021practice}. These dependencies have traditionally made RSSI an imperfect observable for ranging. However, when the link geometry is controlled or calibrated, the perturbation caused by an intervening object can itself become a useful sensing feature. For example, human-body shadowing has been shown to cause significant, channel-dependent changes in BLE RSSI \cite{naghdi2019trilateration, naghdi2020detecting}.

This work investigates a deliberately constrained question: \emph{Can ordinary RSSI measurements from commodity BLE devices distinguish broad classes of common line-of-sight obstructions without requiring an additional sensor?} To answer this, we construct a two-node testbed using nRF5340 development kits and collect both an unobstructed distance baseline and fixed-geometry obstruction measurements. The main contributions of this paper are as follows:
\begin{itemize}
    \item We present a low-cost sensing architecture that relies solely on standard BLE transmissions and receiver-reported RSSI measurements using commodity nRF5340 hardware.
    \item We separate distance-dependent path loss from obstruction-induced loss. Specifically, a path-loss exponent is estimated from measurements collected over multiple distances, while an object observed at a fixed distance is characterized by its excess attenuation relative to a matched unobstructed baseline.
    \item We report preliminary measurements for common dielectric, biological, and conductive obstructions and analyze their separability and variability.
    \item We formulate a lightweight classification and evaluation pipeline based on summary statistics of an RSSI time series, making it suitable for future on-device implementation.
\end{itemize}

The current measurements establish the feasibility of the proposed approach rather than claiming environment-independent material recognition. Accordingly, we explicitly describe the additional validation required to quantify classification accuracy across unseen trials, distances, and environments.

\section{Related Work}
Material identification using a variety of sensing methods is a well-studied topic in the literature~\cite{feng2019wimi,gu2021wimate,agresti2019material,tellbach2023wireless}. It has been investigated using commodity Wi-Fi devices in~\cite{feng2019wimi} and commercial Wi-Fi devices in~\cite{gu2021wimate}. Material identification using RF sensors and convolutional neural networks was studied in~\cite{agresti2019material}, while a chipless RFID tag-based approach was presented in~\cite{tellbach2023wireless}.

Material identification using wireless sensing can broadly rely on reflected, transmitted, or scattered signals. Reflection-based approaches infer the dielectric properties of a material from the returned signal, whereas penetration-based approaches measure the attenuation and phase changes of a signal propagating through the material \cite{chen2025survey}. Radar and UWB systems provide fine time or frequency resolution, while RFID-based systems can exploit tag response characteristics. Although these capabilities improve material discrimination, they often require dedicated hardware or specialized deployment configurations. In contrast, the present work investigates a penetration-based observable---packet RSSI---that is readily available from a commodity BLE receiver.

\subsection{BLE RSSI Characterization}
BLE RSSI has been extensively studied for proximity detection~\cite{montanari2017study,ng2019compressive,sachan2024proximity}, indoor positioning~\cite{ramirez2021practice,jianyong2014rssi}, outdoor positioning~\cite{shin2022outdoor}, and vehicle tracking~\cite{gil2026vehicle}. Given the widespread use of RSSI measurements in BLE-based systems, the authors of~\cite{yadav2026design} experimentally evaluated the performance of BLE-based IoT systems. The authors of~\cite{ramirez2021practice} demonstrated that antenna orientation, environmental conditions, and filtering significantly affect RSSI-based distance estimation. Similarly, the authors of~\cite{naghdi2019trilateration} modeled human-body shadowing and estimated an online path-loss exponent to improve BLE positioning. Their subsequent work~\cite{naghdi2020detecting} further showed that the effects of obstacles can vary across the three BLE advertising channels. In these studies, obstruction-induced variations are primarily treated as sources of error to be mitigated. In contrast, our objective is complementary: after distance calibration, the obstruction-induced component is treated as the sensing target.

\subsection{Research Gap}
Existing high-resolution material-sensing techniques generally rely on specialized RF measurements, whereas studies using commodity BLE devices primarily employ RSSI for ranging or localization. Consequently, there is limited evidence regarding the extent to which a single RSSI stream from an unmodified embedded BLE pair can support broad obstruction identification. This work addresses that gap while avoiding the stronger, and currently unsupported, claim that a single RSSI value can uniquely identify a material.

\section{System Model and Sensing Method}
\subsection{BLE Testbed}
The experimental testbed consists of two nRF5340 development kits, denoted by $\mathrm{Tx}$ and $\mathrm{Rx}$. The nRF5340 is a dual-core Bluetooth 5.3 system-on-chip equipped with a 2.4-GHz radio~\cite{nordicnrf5340}. The transmitter periodically sends BLE packets at a fixed transmit-power setting, while the receiver records the RSSI reported for each received packet and forwards the timestamped measurements to a host computer via USB--UART. No external RF sensor, antenna array, or modification of the BLE physical layer is required.

\begin{figure}[t]
    \centering
    \includegraphics[width=0.98\columnwidth]
    {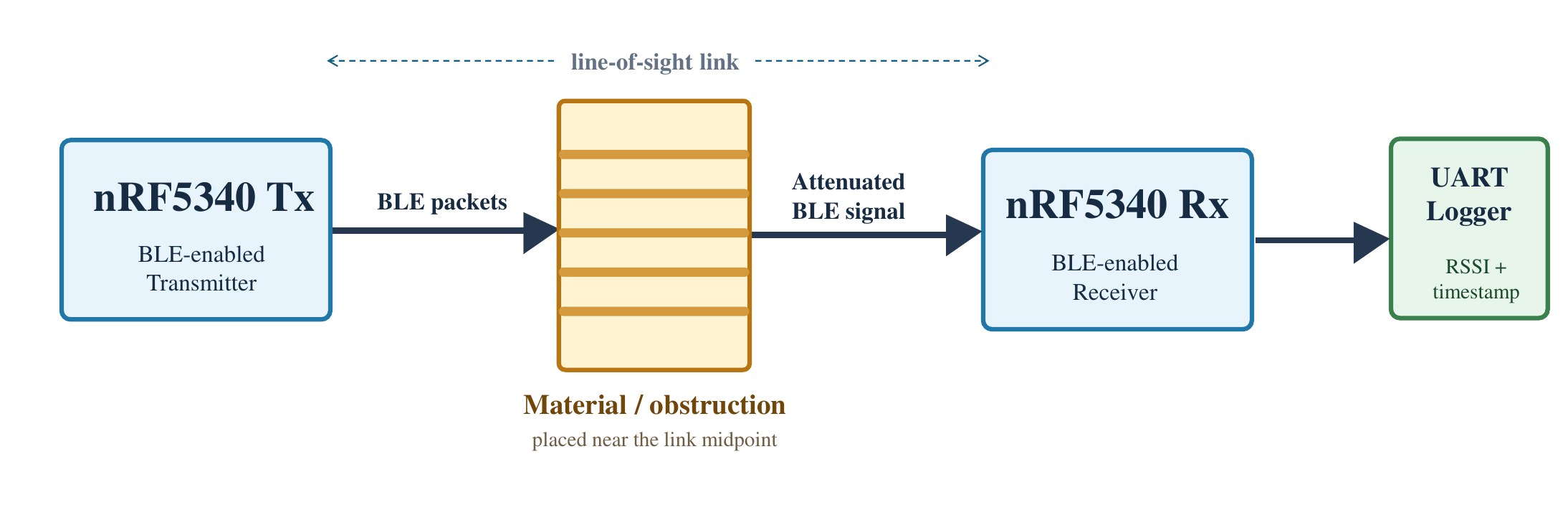}
    \caption{Commodity-BLE material-sensing testbed. The object under test is placed near the midpoint of the line-of-sight link.}
    \label{fig:testbed}
\end{figure}

For obstruction measurements, the boards are kept stationary, and the material sample is placed approximately perpendicular to the line of sight. The experimental configuration is illustrated in Fig.~\ref{fig:testbed}. The primary observable is the RSSI sequence
\begin{equation}
    \mathcal{R}_{m,d}=\{r_{m,d}[1],r_{m,d}[2],\ldots,r_{m,d}[N]\},
\end{equation}
where $m$ represents the obstruction class, $d$ denotes the link distance, and $N$ is the number of received packets retained for each trial.

\begin{figure}[t]
    \centering
    \includegraphics[width=0.92\columnwidth]
    {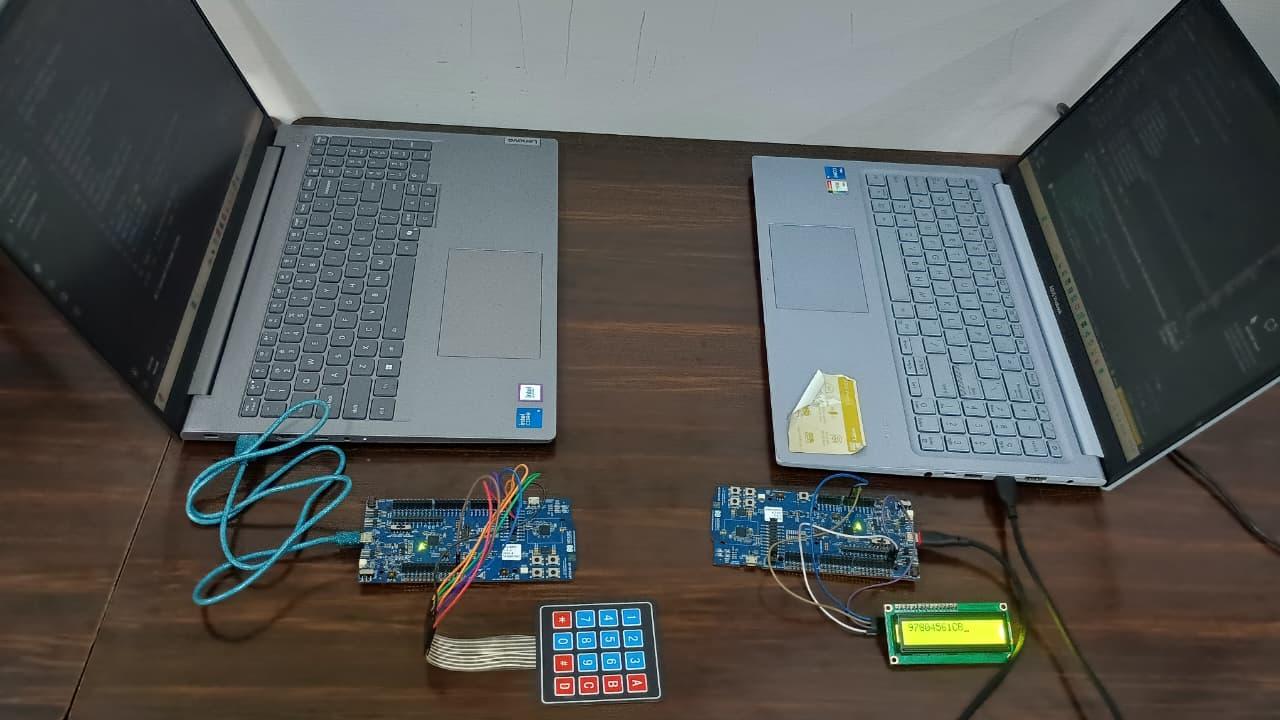}
    \caption{The complete BLE experimental prototype consists of two nRF5340 DK nodes, host computers for firmware programming and UART logging, a matrix keypad connected to the peripheral node, and a character LCD connected to the receiver.}
    \label{fig:complete_testbed}
\end{figure}

Figure~\ref{fig:complete_testbed} shows the implemented bench prototype used to verify BLE packet exchange and RSSI acquisition. The host computers are used for firmware programming and serial logging, while the keypad and LCD demonstrate that material sensing can coexist with conventional BLE application functions. These components are not additional RF-sensing devices and are not required for the proposed sensing principle.

\subsection{Distance Calibration}
For a log-normal shadowing model, the RSSI at distance $d$ can be expressed as
\begin{equation}
 r_0(d)=A-10\alpha\log_{10}\left(\frac{d}{d_0}\right)+X_\sigma,
 \label{eq:pathloss}
\end{equation}
where $A$ represents the mean RSSI at the reference distance $d_0$, $\alpha$ denotes the path-loss exponent, and $X_\sigma$ is a zero-mean Gaussian shadowing term with standard deviation $\sigma$. Importantly, the baseline RSSI varies with distance. Given controlled packet-level unobstructed measurements $\bar r_0(d_k)$, the parameters $A$ and $\alpha$ can be estimated using least-squares regression:
\begin{equation}
 (\hat A,\hat {\alpha})=\arg\min_{A,\alpha}\sum_k
 \left[\bar r_0(d_k)-A+10\alpha\log_{10}(d_k/d_0)\right]^2.
 \label{eq:fit}
\end{equation}
In this study, the 1--5~m traces are experimentally recorded using the testbed shown in Fig.~\ref{fig:complete_testbed}.

\subsection{Excess Attenuation Caused by an Obstruction}
Assigning material identity from a single absolute RSSI level conflates material-induced attenuation with propagation distance. Instead, we define the mean excess attenuation of material $m$ at distance $d$ as
\begin{equation}
 L_{\mathrm{ex}}(m,d)=
 \bar r_0(d)-\bar r_m(d),
 \label{eq:excess}
\end{equation}
where $\bar r_0(d)$ and $\bar r_m(d)$ denote the mean RSSI values for a temporally matched unobstructed trial and a material trial, respectively. A positive value of $L_{\mathrm{ex}}$ indicates additional attenuation introduced by the material. The session-matched subtraction reduces offsets caused by distance, transmit power, receiver calibration, and slow environmental drift.

\subsection{Feature Extraction and Classification}
For each observation window, we define the candidate feature vector
\begin{equation}
 \mathbf{x}=[L_{\mathrm{ex}},\,s_r,\,\mathrm{IQR}_r,\,
 q_{10},\,q_{90},\,\rho_1,\,p_{\mathrm{loss}}]^\mathsf{T},
 \label{eq:features}
\end{equation}
where $s_r$ denotes the RSSI standard deviation, $\mathrm{IQR}_r$ represents the interquartile range, $q_{10}$ and $q_{90}$ are the 10th and 90th percentiles, $\rho_1$ is the lag-one autocorrelation, and $p_{\mathrm{loss}}$ is the packet-loss rate. A lightweight classifier $g(\mathbf{x})$ maps the feature vector to an obstruction label. In this work, a simple threshold-based classifier is employed, while more advanced classifiers, such as $k$-nearest neighbors, support-vector machines, and decision trees, can be explored in future studies.

\section{Experimental Methodology}

\subsection{Data Sources and Provenance}
\begin{figure}[t]
    \centering
    \includegraphics[width=0.95\columnwidth]
    {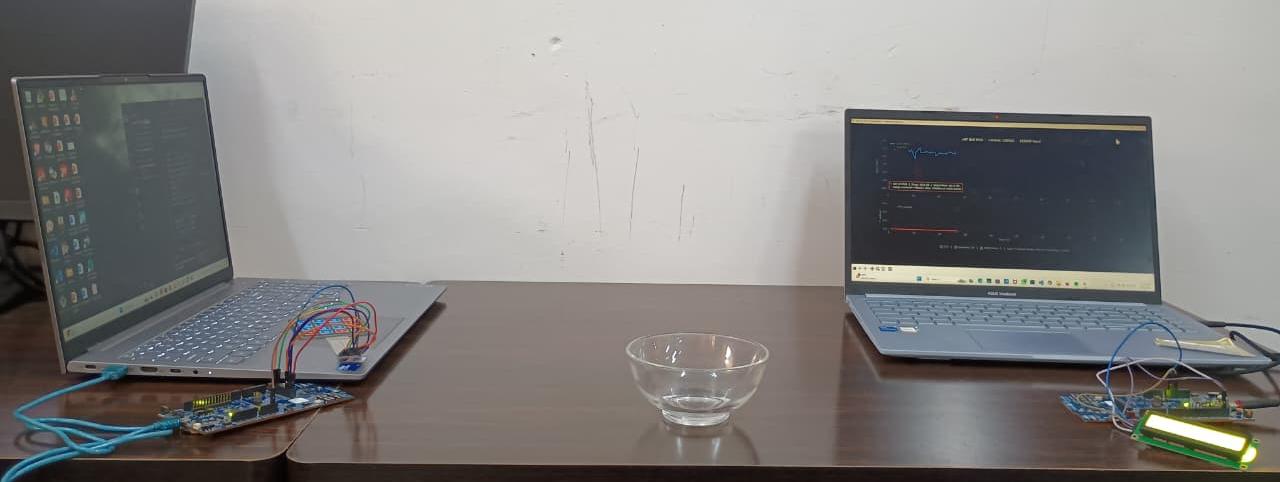}
    \caption{Experimental arrangement for the glass-obstruction trial. The
    glass object is placed near the midpoint of the line-of-sight path between the nRF5340 transmitter and receiver while RSSI is recorded through the receiver's UART logging interface.}
    \label{fig:glass_obstruction}
\end{figure}

Figure~\ref{fig:glass_obstruction} illustrates the geometry of a representative material trial. The transmitter and receiver remain stationary, and only the glass object is introduced into the propagation path. Maintaining fixed radio positions helps associate the resulting temporal RSSI perturbation with the obstruction rather than with changes in link distance.\footnote{Future studies should record the exact glass thickness, orientation, and placement tolerances to improve identification accuracy and experimental reproducibility.}

During each material trial, the transmitter and receiver remain stationary, while the object under test is positioned near the midpoint of the line-of-sight path, as illustrated for glass in Fig.~\ref{fig:glass_obstruction}. This arrangement is intended to associate temporal RSSI variations with the introduced object. However, the available records do not consistently document material thickness, exact orientation, BLE advertising channel, or repeated measurements.

\begin{figure*}[t]
    \centering
    \includegraphics[width=0.92\textwidth]
    {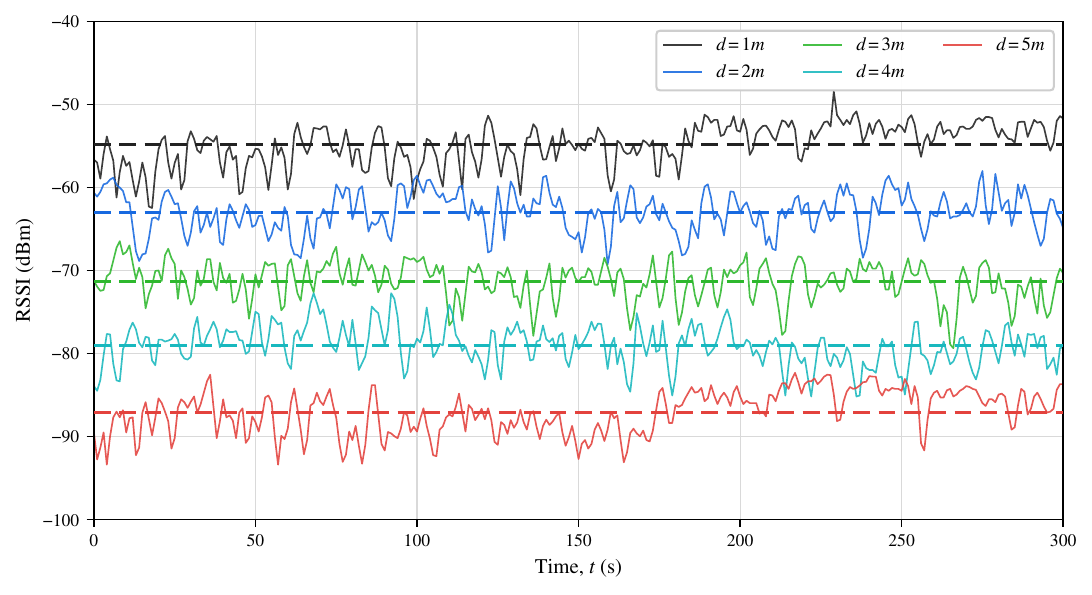}
    \caption{RSSI variation over time for transmitter--receiver separations from 1 to 5~m. Solid curves show the traces, while the color-matched dashed horizontal lines show their temporal means. The data are used for material classification.}
    \label{fig:distance_calibration}
\end{figure*}

The distance plot in Fig.~\ref{fig:distance_calibration} presents RSSI traces measured over 0--300~s for transmitter--receiver separations of 1, 2, 3, 4, and 5~m. These original packet-level traces are used as unobstructed RSSI baselines for material identification.

\subsection{Analysis Procedure}
For each distance trace, we compute only its temporal mean and inspect the spread about that mean. The resulting trend is used as calibration context: an absolute RSSI level cannot be interpreted as a material signature without accounting for link distance. 

For a recorded material sequence, let $r_m[t]$ denote the RSSI value and $b_m[t]$ denote its local moving average. We characterize a downward transient using
\begin{equation}
    \delta_m[t]=b_m[t]-r_m[t].
    \label{eq:transient_deviation}
\end{equation}
A positive value of $\delta_m[t]$ indicates that the instantaneous RSSI is below its local average. Unlike $L_{\mathrm{ex}}$ in \eqref{eq:excess}, this quantity is not referenced to an independently acquired, unobstructed trial. Therefore, it is used to compare candidate temporal behaviors rather than as a calibrated material attenuation coefficient.

\section{Experimental Results and Discussion}
In our experiments, five unknown material samples were considered. Each sample was individually placed in the line-of-sight (LoS) path between the transmitter ($\mathrm{Tx}$) and receiver ($\mathrm{Rx}$) while maintaining a constant $\mathrm{Tx}$--$\mathrm{Rx}$ separation. The obstructing material was positioned at the near midpoint of the link, and the measured RSSI values were used to estimate $\hat{\alpha}$ using~\eqref{eq:fit}. The material was then classified by comparing the estimate $\hat{\alpha}$ with the trained dataset. Since the training dataset contained only a limited number of known materials, the classifier correctly identified samples belonging to the trained classes but misclassified unseen materials. As shown in the last column of Table~\ref{tab:material_reconstructed}, human body, glass, and wood samples were correctly identified. This demonstrates the effectiveness of the proposed approach in achieving higher classification accuracy.  However, brass and paper, which were not included in the training set, were classified as ceramic and a plastic bottle, respectively. These results demonstrate that classification performance is limited by the diversity of the training dataset. Therefore, improving accuracy and generalization requires a larger and more diverse collection of training materials.

\subsection{Distance Dependence as Calibration Context}

Figure~\ref{fig:distance_calibration} shows a progressive reduction in mean RSSI with increasing distance $d$. The approximate mean RSSI values at 1, 2, 3, 4, and 5~m are $-54.85$, $-62.99$, $-71.34$, $-79.05$, and $-87.06$~dBm, respectively. Relative to the 1~m measurement, the corresponding reductions are 8.14, 16.49, 24.20, and 32.22~dB, resulting in an average reduction of approximately $8$~dB per metre over the considered range. However, instantaneous RSSI samples fluctuate around each mean due to multipath propagation, interference, and environmental motion. Therefore, the distance characterization is not itself the sensing objective; rather, it establishes that material identification should rely on distance-matched baselines or baseline-relative features instead of a single global RSSI threshold.

\subsection{Material-Specific RSSI Behavior}
\begin{figure*}[t]
    \centering
    \includegraphics[width=0.96\textwidth]
    {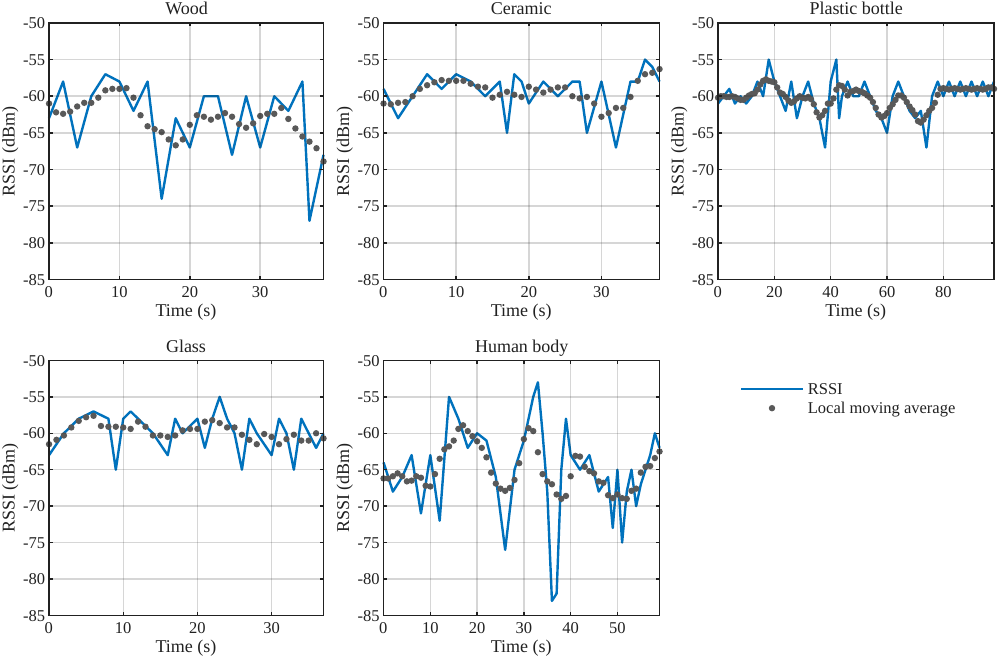}
    \caption{RSSI traces for wood, ceramic, plastic bottle, glass, and a human-body obstruction. Solid curves denote RSSI, and dotted curves denote local moving averages computed from the same traces.}
    \label{fig:material_traces}
\end{figure*}

To train the classifier model $g(\mathbf{x})$, the known sample is placed near the midpoint of the LoS path between the $\mathrm{Tx}$ and $\mathrm{Rx}$, and the resulting RSSI trace is recorded, as shown in Fig.~\ref{fig:material_traces}. The experiment is repeated at a fixed $\mathrm{Tx}$--$\mathrm{Rx}$ separation for five material objects: wood, ceramic, an empty plastic bottle, glass, and a human body. Figure~\ref{fig:material_traces} presents the five material sequences using identical vertical scales. The solid curves represent the RSSI measurements, while the dashed curves represent the corresponding local averages.

\begin{table}[t]
\centering
\caption{Statistics of the Known Material Traces}
\label{tab:material_reconstructed}
\resizebox{\columnwidth}{!}{%
\begin{tabular}{lcllcl}
\toprule
Material & Samples & Mean & SD &  Peak $\delta_m[t]$ & $\hat{\alpha}$ \\
 & & (dBm) & (dB) & (dB) & \\
\midrule
Wood             & 40 & $-63.03$ & 4.62 &  10.8 & 3.20 \\
Ceramic  & 39 & $-59.42$ & 2.54 &   5.6 & 2.70 \\
Plastic bottle   & 99 & $-60.14$ & 2.17 &   4.5 & 2.30  \\
Glass            & 38 & $-59.78$ & 2.42 &   5.9 & 2.50  \\
Human body       & 60 & $-65.00$ & 5.71 &  16.0 & 4.24  \\
\bottomrule
\end{tabular}}
\end{table}

Table~\ref{tab:material_reconstructed} summarizes the known material features. Wood and Human body exhibit a peak transient deviation $\delta_m[t]$ of 10.8~dB and 16.0~dB, respectively. Other materials ceramic, plastic, and glass remain closer to their local averages, with peak deviations of 5.6, 4.5, and 5.9~dB, respectively. 
These differences support coarse material grouping, particularly the separation of strong biological or bulky obstructions from lower-perturbation dielectric objects. Therefore, the table represents candidate features and also provides estimated $\hat{\alpha}$ values computed using~\eqref{eq:fit} and they are found to be close to their actual $\alpha$ values.

\subsection{Identification of Unknown Obstructions}
\begin{table}
\centering
\caption{Statistics of the Unknown Material Traces}
\label{tab:material_unknown}
\resizebox{\columnwidth}{!}{%
\begin{tabular}{lcllcll}
\toprule
Material & Samples & Mean & SD &  Peak $\delta_m[t]$ & $\hat{\alpha}$ & Identified Material \\
 & & (dBm) & (dB) & (dB) & & (Actual Material)\\
\midrule
Unknown1       & 50 & $-48.60$ & 4.30 &  4.1 & 2.21 & Plastic bottle (Paper)\\
Unknown2      & 56 & $-50.80$ & 7.90 &  18.9 & 4.80 & Human body (Human body)\\
Unknown3       & 62 & $-48.60$ & 2.71 &  9.1 & 2.74 & Ceramic (Brass)\\
Unknown4      & 48 & $-58.41$ & 2.30 &  5.8 & 2.42 & Glass (Glass) \\
Unknown5      & 44 & $-49.80$ & 4.30 &  10.7 & 3.10 & Wood (Wood) \\
\bottomrule
\end{tabular}}
\end{table}

To identify the unknown materials, we placed them one at a time at the midpoint between the $\mathrm{Tx}$ and $\mathrm{Rx}$, following the material-sensing testbed shown in Fig.~\ref{fig:testbed}. We recorded the corresponding RSSI traces and computed the statistical features reported in Table~\ref{tab:material_unknown}. Based on the estimated path-loss exponent $\hat{\alpha}$, the classifier $g(\mathbf{x})$ identifies each material. The classification results are presented in the last column of Table~\ref{tab:material_unknown}, together with the actual obstructing materials indicated in parentheses. The results show that $g(\mathbf{x})$ correctly identified the human body, glass, and wood. However, paper and brass were misclassified as a plastic bottle and ceramic, respectively. These errors occurred because paper and brass were not represented in the training dataset, although their $\hat{\alpha}$ values were estimated reasonably accurately.

\subsection{Application Context}
Material identification can augment an existing BLE link without requiring an additional RF sensor. Potential applications include detecting a human entering a monitored path, distinguishing a low-loss object from a strong obstruction, identifying blockage of a storage compartment, or adapting communication parameters when a material alters the link characteristics. The keypad, LCD, and temperature-transfer functions in the prototype serve as application demonstrations; the primary material-sensing contribution is the extraction of obstruction-related information from RSSI measurements already generated by the communication radio.

\subsection{Limitations}
The study uses a single transmitter--receiver pair and an incompletely documented indoor geometry. Material thickness, area, moisture content, orientation, BLE channel, transmit power, and co-channel interference were not consistently recorded. Furthermore, only five reconstructed sessions represent each material, which may confound material identity with session-specific conditions and link geometry.

\section{Conclusions and Future Work}
\label{sec:conclusion}
This paper investigated device-free material and obstruction identification using RSSI measurements from commodity nRF5340 BLE radios. The distance-calibration experiment showed that the mean RSSI decreased from approximately $-54.85$~dBm at 1~m to $-87.06$~dBm at 5~m, confirming that absolute RSSI alone cannot provide a distance-independent material signature. Consequently, RSSI statistics, transient deviation, and the estimated path-loss exponent were considered as candidate sensing features. Among the trained material classes, the human body produced the largest RSSI variability and transient deviation, whereas ceramic, plastic, and glass produced smaller deviations from their respective local baselines.

The threshold-based classifier correctly identified the human body, glass, and wood in the unknown-material trials. However, brass and paper  were misclassified as a ceramic and plastic bottle, respectively, because these materials were not represented in the training dataset. The results establish the feasibility of low-cost, device-free material identification using commodity BLE RSSI, while highlighting the limited ability of the current approach to recognize unseen materials.

Future research can extend this study by developing a more diverse dataset comprising dielectric, biological, and conductive materials with variations in thickness, orientation, moisture content, transmitter--receiver distance, and placement. Researchers should conduct multiple independent trials for each material--distance configuration using a matched unobstructed baseline and recording the corresponding BLE-channel information. Future studies may also compare the threshold-based classifier with advanced methods such as $k$-nearest neighbors, support-vector machines, decision trees, and lightweight ensemble models. Further investigation should consider on-device implementation and evaluation using different nRF5340 pairs and indoor environments to quantify computational cost, robustness, reproducibility, and generalization performance.

\bibliographystyle{ieeetr}
\balance
\bibliography{references.bib}

\end{document}